\documentclass[10pt,a4paper,twocolumn,english,showpacs]{revtex4}

\usepackage{graphicx}
\usepackage{epsfig}
\usepackage[english]{babel}
\usepackage{amsmath}
\usepackage{amssymb}
\usepackage{amsfonts}
\usepackage{longtable}
\usepackage{dcolumn}% Align table columns on decimal point
\usepackage{bm}

\newcommand{\be}{\begin{equation}}
\newcommand{\bea}{\begin{eqnarray}}
\newcommand{\eea}{\end{eqnarray}}
\newcommand{\ee}{\end{equation}}

\begin{document}
\title{DMRG Simulation of the $\text{SU}(3)$ AFM Heisenberg Model}

\author{M. Aguado}
\affiliation{Max-Planck-Institut f\"ur Quantenoptik.  Hans-Kopfermann-Str. 1.  D-85748 Garching,
Germany}
\author{M. Asorey}
\affiliation{Departamento de F\'{\i}sica Te\'{o}rica, Facultad de Ciencias Universidad de Zaragoza,
50009 Zaragoza, Spain}
\author{E. Ercolessi, F. Ortolani}
\affiliation{Dipartimento di Fisica dell'Universit\`{a} di Bologna and INFN, 
Via Irnerio 46, 40127 Bologna, Italy} 
\author{S. Pasini}
\email{pasini@fkt.physik.uni-dortmund.de}
\affiliation{Lehrstuhl f\"{u}r Theoretische Physik I, Universit\"{a}t Dortmund, Otto-Hahn Stra\ss{}e 4, 44221 Dortmund, Germany}

\date{\rm\today}

\begin{abstract}
We analyze the antiferromagnetic $\text{SU}(3)$ Heisenberg chain by means of the
Density Matrix Renormalization Group (DMRG). The results confirm that the
model is critical and the computation of its central charge and the 
scaling dimensions of the first excited states show that the underlying
low energy  conformal field theory is  the $\text{SU}(3)_1$ Wess-Zumino-Novikov-Witten 
model.
\end{abstract}


\maketitle

\section{Introduction\label{s1_intro}}
In recent years, a renewed interest in models of condensed matter with
a symmetry larger than SU(2) has arisen. This is because
these models represent not only challenging theoretical problems but
also  can be effectively implemented experimentally. In
particular SU(4) systems can be realized in laboratories in
transition metals oxides \cite{tokura} where the electron spin is
coupled to the orbital degrees of freedom.  A
possible realization  of  SU(3) antiferromagnetic (AFM) spin chains
in systems of ultracold atoms in optical lattices
has been recently proposed \cite{greiter}. In this
case the spin would be related to the SU(3) rotation in an
internal space spanned by the three available  atomic states 
(colors, in the SU(3) language), with the condition that the
number of particles of each color is conserved. Other examples
involve the SU(3) trimer state  in a spin
tetrahedron chain \cite{chen05,chen06}, or the spin tube models in
a magnetic field \cite{orignac} where the low-energy effective
Hamiltonian can be identified with a particular anisotropic SU(3)
spin chain.\\
\indent
From a theoretical point of view, the SU(3) spin model has also
been studied from different viewpoints. In recent years the interest
on ferromagnetic SU(N) spin chains has been boosted by
their implication in the AdS/CFT correspondence \cite{zarembo,beisert}.
On the other side the family of integrable spin chains include some
models with SU(3) symmetry,  as first shown by
 Sutherland \cite{suth}, who generalized the Bethe-Ansatz 
to multiple component systems which include the SU(3) spin chain,
showing that it is gapless. Also the SU(3) Heisenberg model can be
directly related to a particular SU(3)-symmetric  bilinear
biquadratic spin-1 chain, the Lai-Sutherland (LS) model, which is
also known to be critical \cite{lai_sut,chang_affl}. In terms of 
Conformal Field Theory (CFT) the LS model
and the SU(3) spin chain should belong to the same universality class, 
that of the  $\text{SU}(3)_1$ Wess-Zumino-Novikov-Witten (WZNW)
model \cite{affl86,affl88}.

In this paper we present a numerical analysis of
the SU(3) spin chain by means of the Density Matrix Renormalization
Group (DMRG). After a short description of the model and its mathematical 
framework (Section \ref{s2_model}), we  present  our new results
 (Section \ref{s3_NumAn}) which confirm the criticality of
the model as well as its correspondence to the Lai-Sutherland
model. In particular, due to the ability of our program to
provide the quantum numbers for each state, we can show that the excited states
of the spin chain have the same quantum numbers as the irreducible representations (IR) of SU(3).
We compute the scaling dimensions of the first
excitations which turn out to agree with those of the $\text{SU}(3)_1$ WZNW model 
which corresponds to  the low energy effective field theory descriptions 
of our spin chain. The results are further confirmed 
by the computation of the central charge by means of the 
vacuum entanglement entropy.

\section{The SU(3) model\label{s2_model}}
We consider the following Heisenberg model 

\be\label{hamilt_1}H=J\sum_{i=1}^L \mathbf{S}_i\cdot\mathbf{S}_{i+1}\ee where the
spin variables are expressed in terms of the generators of 
$\text{SU}(3)$ in the fundamental representation:
$\mathbf{S}^a_i=\frac{1}{2}\lambda_i^a$, with
$a=1,..,8$ and $\lambda_a$ the eight Gell-Mann matrices. The sign of $J$ selects
respectively an antiferromagnetic spin chain ($J>0$) or a
ferromagnetic (FM) one ($J<0$). In the following we shall
concentrate only on the AFM case, which has been partially considered also in ref.
\cite{itoi,fue}.\\
In terms of the  following ladder operators, $T^\pm=\lambda^1\pm
i\lambda^2$, $V^\pm=\lambda^4\pm i\lambda^5$ and $U^\pm=\lambda^6\pm
i\lambda^7$, the Hamiltonian (\ref{hamilt_1}) becomes

\bea\label{hamilt_3}\nonumber
H&=&\frac{J}{2}\sum_{i=1}^L\left\{\frac{1}{4}\left(T^+_iT^-_{i+1}+
V^+_iV^-_{i+1}+U^+_iU^-_{i+1}+ h.c.\right)\right.
\\&&+ \left.\frac{1}{2}\lambda_i^3\lambda_{i+1}^3
+\frac{1}{2}\lambda_i^8\lambda_{i+1}^8\right\}.\eea This makes 
easier to identify two operators, $S_z$ and $Q_z$, given by the sums of the
two diagonal Gell-Mann matrices 
\bea\label{quantnumb}S_z=\sum_i\frac{1}{2}\lambda_i^3 \qquad
Q_z=\sum_i\frac{\sqrt{3}}{2}\lambda_i^8,\eea that commute with the
Hamiltonian and correspond to conserved quantities (isospin and hypercharge). 
The corresponding  quantum numbers  label  the different eigenstates of (\ref{hamilt_1}).

The Lai-Sutherland model  is defined as
the bilinear biquadratic spin-1 chain

\be\label{biquad}H=J'\sum_{i=1}^L\left[\tilde{\mathbf{S}}_i\cdot\tilde{\mathbf{S}}_{i+1}+
(\tilde{\mathbf{S}}_i\cdot\tilde{\mathbf{S}}_{i+1})^2\right],\ee
and characterized by an $\text{SU}(3)$ symmetry. The model (\ref{biquad}) and the SU(3) spin chain
can be mapped one onto the other by
means of the following identity \cite{schmitt}

\be\label{SU3vsBiQuad} \tilde{{\mathbf S}}_i\tilde{{\mathbf
S}}_{i+1}+(\tilde{{\mathbf S}}_{i}\tilde{{\mathbf
S}}_{i+1})^2-1=\frac{1}{3}+\frac{1}{2}\sum_{a=1}^8\lambda^a_{i}\lambda^a_{i+1}
.\ee We have already mentioned in the introduction that the LS
model is known to be gapless and to belong to the same
universality class of the SU(3) level-1 Wess-Zumino-Novikov-Witten
model with central charge $c=2$. Due to the
correspondence between the two models, the $\text{SU}(3)_1$ WZNW model has to be
the low energy effective critical field theory also for the SU(3) spin chain. 
We shall numerically show that the SU(3) Heisenberg chain is
critical, and from the energy state obtained from the DMRG,
we shall  compute the central charge and the scaling
dimensions of (\ref{hamilt_1}) and compare them to
the values predicted for the $\text{SU}(3)_1$ WZNW model.

The states of the spin chain can be 
organized according to the irreducible representations of the affine (Kac-Moody) Lie algebra associated to SU(3). Let us recall \cite{diFrancesco} that a useful way of representing the IR's of the Lie algebra $su(3)$ is through the Young Tableau (YT) which can be labelled by two positive integer numbers $(p,q)$.
Once $p$ and $q$ are known, one can easily compute the dimension $d$ of the representation and the quantum numbers associated to the isospin $S_z$ and  the hypercharge $Q_z$ according to \cite{Mukunda,diFrancesco}:
\begin{equation}
 \label{dimIR_DiFrancesco}
d= \frac{1}{2}(p+1)(q+1)(p+q+2)
\end{equation} 
and
\begin{equation}
\label{quantum numbers}  \begin{array}{l}  S_z= -I, -I+1, ..., I-1,I \\ ~ \\ Q_z= \frac{3}{2} Y \end{array}
 \end{equation}
 where $  I=\frac{1}{2}(r+s)$ and $Y = (r-s) - \frac{2}{3} (p-q)$, with $0\leq r\leq p \, , \, 0\leq s\leq q$ .
In particular, the cases $(1,0)$ and $(0,1)$ give respectively the fundamental ($\mathbf{3}$) and the anti-fundamental ($\mathbf{\bar{3}}$) IR, while the singlet representation ($\mathbf{1}$) corresponds to $(0,0)$. 

It has been proved \cite{hakobyan} that, in analogy with the SU(2) case, the ground state (GS) of the AFM SU(3) Hamiltonian is a singlet and, since  it is made of particles $u$, $d$ and $s$ in equal number, it can be obtained in finite chains having only a number of sites which is a multiple of three, $L=3M$. As for the excited states, we expect them to be in correspondence with the tower of conformal states of the corresponding SU(3) WZNW model. The primary states of this theory are a finite number and are given \cite{diFrancesco} by fields $\Phi_{\lambda, \bar{\lambda}}$, whose holomorphic (antiholomorphic) part transforms according to a representation $\lambda=(p,q)$ ($\bar{\lambda}=(p',q)$) with the values of $p,q$ (and similarly of $p',q'$) satisfying the condition: $p+q\leq k$, $k$ being the level. The conformal dimension of the primary field is then $x_{\lambda, \bar{\lambda}} = x_{(p,q)} + x_{(p',q')}$ with
\begin{equation}
 \label{xj_DiFrancesco_finale}
x_{(p,q)}=\frac{1}{3(k+N)}(p^2+q^2+pq+3p+3q),
\end{equation} 
and a similar expression for $x_{(p',q')}$. For future reference, the values of $x_{\lambda, \bar{\lambda}}$ for some primary fields in the case of $k=1$ are reported in Table \ref{IR_quantum_numb}.

%%%%%%%%%%%%%%%%%%%%%%%%%%%%%%%table%%%%%%%%%%%%%%%%%%%%%%%%%%%%%%%%%%%%%
\begin{table}[t] 
\begin{center}
\begin{tabular}{|c|c|c|c|c|}
\hline  
  $\lambda$ &  $\bar{\lambda}$ & \phantom{\Large{I}} $(S_z, Q_z)$ & $x_{\lambda, \bar{\lambda}}$\phantom{\Large{I}} \\
\hline\hline

 ($\mathbf{1}$)  &  ($\mathbf{1}$)  & (0,0) & 0\\
($\mathbf{3}$)  &  ($\mathbf{1}$)    & $\left\{ \begin{array}{cc} (\pm 1/2,1/2) \\ (0,-1)\end{array} \right.$ & 1/3\\ 
 ($\mathbf{\bar{3}}$) &  ($\mathbf{1}$)     & $\left\{ \begin{array}{cc} (\pm 1/2,-1/2) \\ (0,1)\end{array}  \right.$ & 1/3\\
($\mathbf{3}$) & ($\mathbf{\bar{3}}$)  &  $\left\{ \begin{array}{cc} (\pm 1/2,\pm 3/2) \\ (0,0) \mbox{ \small{(3 times)} }\\ (\pm 1,0)   \end{array}  \right.$ & 2/3 \\
\hline
\end{tabular}
\caption{Quantum numbers and scaling dimensions for some of the  primary fields $\Phi_{\lambda, \bar{\lambda}}$ of the SU(3)$_1$ WZNW model.
 \label{IR_quantum_numb}}
\end{center}
\end{table}
%%%%%%%%%%%%%%%%%%%%%%%%%%%%%%%%%%%%%%%%%%%%%%%%%%%%%%%%%%%%%%%%%%%%%%% 

To end up this section, we notice that in a finite chain of length $L$ not all quantum numbers, i.e. states, may be realized. For examples, working with periodic boundary conditions and with an even number of sites, the singlet ($\mathbf{1})\times( \mathbf{1})$ (ground) state, with $x=0$, appears only for chains with $L=6M$ (with $M$ a positive and integer number), while the  $(\mathbf{3})\times (\mathbf{1})$ (or the $(\mathbf{1})\times (\mathbf{\bar{3}})$) states are present only if $L=6M+4$ (or $L=6M+2$), both with $x=1/3$.

\section{Numerical analysis \label{s3_NumAn}}
The SU(3) version of the DMRG we have used implements the
following Hamiltonian 
\begin{eqnarray}
 H&=&\frac{J}{2}\sum_{i=1}^L\left[\frac{1}{4}\left(k_0 \ T^+_iT^-_{i+1}+ k_1 \
V^+_iV^-_{i+1}\right. \right.  \label{hamilt_kdelta}\\ 
&+&k_2\ U^+_iU^-_{i+1}+\left.\left.h.c.\right) +\frac{1}{2}\left(z_0\
\lambda_i^3\lambda_{i+1}^3 +z_1\
\lambda_i^8\lambda_{i+1}^8\right)\right] \nonumber
\end{eqnarray}
 where
$k_j$ and $z_j$ are input parameters. The model
(\ref{hamilt_kdelta}) reproduces the AFM (FM) case when all the $k_j$'s
and the $z_j$'s are equal to 1 (-1). By tuning the input parameters
$k_j$ and $z_j$, we can study all the possible anisotropic version
of the SU(3) Heisenberg model. A very important feature of
this DMRG is that it implements both the quantum numbers $S_z$ and
$Q_z$ given in  (\ref{quantnumb}). This implementation
considerably reduces the computation time and, on the other hand, once
$S_z$ and $Q_z$ are fixed from input, each run of the
DMRG yields exclusively the energies of the states within
those quantum-number sectors. This is very useful when one needs to classify the
excitations according to the values of the
isospin and of the hypercharge.\\
By setting $k_1=k_2=z_1=0$, we restrict to the SU(2) sector of
SU(3). This has been used as a check to the program; the DMRG in
this case reproduces perfectly all the energy states of the SU(2)
Heisenberg model. 

\begin{figure}
\begin{center}
     \includegraphics[width=\columnwidth]{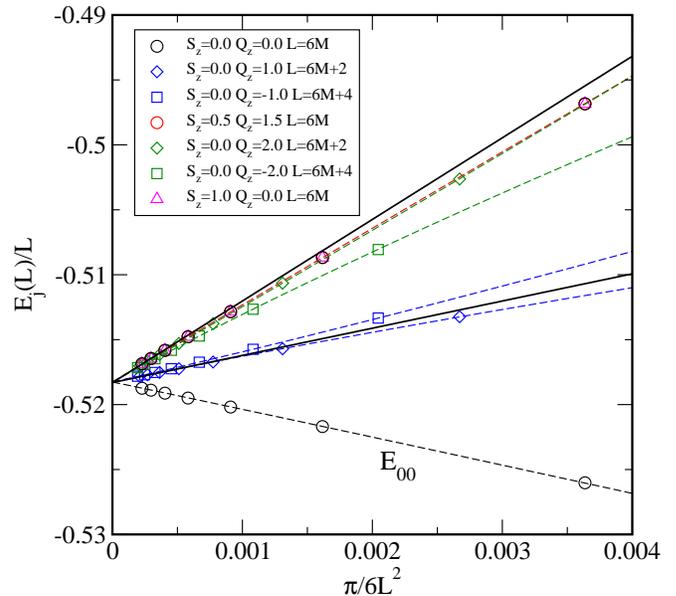}
\end{center}
\caption{Plot of the ground state $E_{00}$ and of the excited states for chains of different lengths (from $L=12$ up to $L=52$). The solid lines have a slope respectively of $\frac{1}{3}$ and $\frac{2}{3}$ and have been drawn as a guide for the eye. For sake of clarity, not all the degeneracies have been reported.} \label{Fig1}
\end{figure}

We study now the isotropic AFM chain with periodic
boundary conditions by means of an infinite size DMRG with up to $m=2200$ states in
order to reduce the uncertainty on the energies to the order of
magnitude of the truncation error.  The
data for the ground state and the first excited states are plotted in Fig. \ref{Fig1}.

%%%%%%%%%%%%%%%%%%%%%%%%%%%%%%%%%%%%%%%%%%%%%%%%%%%%%%%%%%%%
%%%%%%%%%%%%%%%%%%%%%%%%%%%%%%%%%%%%%%%%%%%%%%%%%%%%%%%%%%%%%

Let us first concentrate on the {\it ground state}, which, in agreement with theoretical predictions, it is found only when $L=6M$. The plot of $E_{00}$ as a function of $1/L^2$ shows a good linear behavior; this justifies the fitting of our data by the CFT equations for the GS:
\begin{equation}\label{CFT_GS_1}
\frac{E_{00}}{L}=e_\infty -\frac{\pi c v}{6 L^2}, \end{equation} 
where $e_\infty$ and
the product $cv$ are kept as fitting parameters. We obtain: $e_\infty=-0.518288$ and
$cv=2.04419$. In order to derive the
value of $v$  we need an independent derivation of $c$.
The central charge
for a SU(N) level-k WZNW model  is given by \cite{diFrancesco}
\be\label{centr_ch} c=\frac{k\ (N^2-1)}{k+N}.\ee 
If  the effective
field theory describing our spin chain is the conformal $\text{SU}(3)_1$
WZNW model,  the central charge must be $c=2$.

However, it is possible to have a direct numerical 
derivation of $c$ from the asymptotic behavior of the von Neumann entropy
$S_n=-\mathrm{Tr}_n(\rho_n\log_2\rho_n)$  of  the 
reduced density matrix
 $\rho_n =\mathrm{Tr}_{i>n}\rho$  of a  subchain with $n$ spins   of a critical system of length $L$,
as a function of $n$ and $L$, where $\rho$ is the density matrix associated to the ground state of the chain. Indeed, one has  \cite{holzhey,calabrese}:
\be\label{Sn}S_n=\frac{c}{3}\log_2\left[\frac{L}{\pi}\sin\left(\frac{\pi}{L}n\right)\right]+A.\ee
As usual $c$ is the central charge while $A$ is a non-universal constant. 
%%%%%%%%%%%%%%%%%%%%%%%%%%%%%%%%%%%%%%%%%%%%%%%%%%%%%%%%%%%%
\begin{figure}
\begin{center}
     \includegraphics[width=\columnwidth]{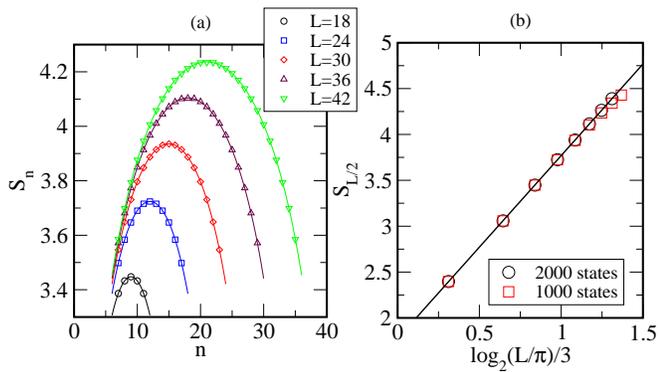}
\end{center}
\caption{Analysis of the von Neumann entropy. The figure on the left (a) shows $S_n$ for $m=1000$ DMRG states as a function of the block length  $n$ for different chains of length $L$; the figure on the right (b) is the plot of the half chain entropy ($n=L/2$) for $m=1000$ and $m=2000$ DMRG states. The linear fit on these data provides the values for the central charge $c$ and the constant $A$.} \label{Fig2}
\end{figure}
%%%%%%%%%%%%%%%%%%%%%%%%%%%%%%%%%%%%%%%%%%%%%%%%%%%%%%%%%%%%%
The DMRG computes the density matrix for a block of length $n$ in a chain of length $L$, so that
$S_n$ becomes quite simple to calculate. Fig. \ref{Fig2} shows the behavior of the von Neumann 
entropy as a function of the block length $n$ in (a) and as a function of the
quantity $y=\log_2\left(\frac{L}{\pi}\right)/3$
(obtained from (\ref{Sn}) by setting $n=L/2$) in (b), for values of the DMRG states equal to 1000 and 2000. 
The data confirm the linear behavior expected from Eq. (\ref{Sn}).
The linear regression $S_n=c y +A$  yields the value for the constant $A=(1.774\pm0.002)$ and for the central charge
$c=(1.995\pm0.001)$. Thus the theoretical prediction of Eq. (\ref{centr_ch}) is confirmed with very high accuracy.  From Fig. \ref{Fig2}(b) it is also evident that the values obtained when keeping only 1000 states in the DMRG run are much less precise. This is the reason why we have then performed all calculations while keeping 2000 states. Finally, the value of $c$  can be substituted into the
 product $cv$ derived from the GS to recover the velocity of the excited modes: $v=(1,0247\pm 0.0005)$, which is close to the expected 
value \cite{suth} $\pi/3$.

Before proceeding with the analysis of the excited states, let us
check  the asymptotic value of the energy density $e_\infty$.
The theoretical prediction for the ground state of the $S=1$ bilinear biquadratic
Heisenberg Hamiltonian (see Ref. \cite{uimin}) is 
\be\label{GS_BiQuad} E_{GS}=-\ln 3-\frac{\pi}{3\sqrt{3}}+1, \ee 
which already takes into account the factor -1 of the l.h.s. of equation (\ref{SU3vsBiQuad}).
Starting from the correspondence between our SU(3) chain and the biquadratic one
(\ref{SU3vsBiQuad}), we can compare the value of $e_\infty$ we
obtained with the one predicted by equation (\ref{GS_BiQuad}):
$E_{GS}=-0.703212$. The match is exact to the third decimal
($-0.703243$), if one also recalls that the Hamiltonian has a factor
$1/4$ (due to the definition of the spin variables in terms of the
SU(3) generators) so that $e_\infty$ needs to be multiplied by a
factor two, and summed to the factor $1/3$ of equation (\ref{SU3vsBiQuad}). 
This is a further numerical proof of the equivalence
between the the Lai-Sutherland and the SU(3) spin model.

Let us study now the {\it excited states}. From  Fig. \ref{Fig1}, one immediately sees that the slope of excited states depends on the length of the chain. In particular, for $L=6M$ the first excitation scales with a slope which is  unmistakably different  from the slope of the $L=6M+2$ or $L=6M+4$ data. For small values of $L$ the data corresponding to the same $S_z$ but with opposite $Q_z$ are split by a finite size correction, while for increasing values of $L$ they tend to overlap and scale to the same asymptotic value. 

For the excited states CFT predicts that:
\be\label{CFT_FistExc} E_j-E_{00}=\frac{2\pi v}{L} x_j \ee 
where
$x_j$ is the scaling dimension of the $j-$th excitation for a given chain of length $L$; $E_{00}$ is given by Eq. (\ref{CFT_GS_1}) where $c$ and $v$ have been derived before and are reported in the caption of Table \ref{tab_scal_dim}. The numerical coefficients for the scaling dimensions that one can obtain from the DMRG data of Fig. \ref{Fig1} are listed in Table \ref{tab_scal_dim}.
%%%%%%%%%%%%%%%%%%%%%%%%%%%%%%%table%%%%%%%%%%%%%%%%%%%%%%%%%%%%%%%%%%%%%
\begin{table}[t]
\begin{center}
\begin{tabular}{|c|c|c|}
\hline  
  $L$   & $x_1$ & $x_2$   \\
\hline  
  6M+2  & $0.3414\pm 0.0001$ & $0.6291\pm 0.0003$\\
  6M+4  & $0.3406\pm 0.0002$ & $0.6238\pm 0.0003$\\
   ($\blacktriangle$)     & $0.3410\pm 0.0002$ & $0.6265\pm 0.0003$\\
\hline
  6M    & - & $0.6503\pm 0.0003$   \\
\hline
\end{tabular}
\caption{Results of the numerical analysis.
The velocity  and the central charge are respectively: $v=(1.0247\pm 0.0005)$ and $c=(1.995\pm0.001)$. The scaling dimensions of the first ($x_1$) and the second ($x_2$) excited states for each $L$ are calculated as described in the text. The mean value $(\blacktriangle)$ between $L=6M+2$ and $L=6M+4$ for $x_1$ and $x_2$ is provided (see also Fig. \ref{Fig1}). For $L=6M$ only the first excitation above the ground state has been considered. \label{tab_scal_dim}}
\end{center}
\end{table}
%%%%%%%%%%%%%%%%%%%%%%%%%%%%%%%%%%%%%%%%%%%%%%%%%%%%%%%%%%%%%%%%%%%%%%%

As expected, the values of the allowed conformal dimensions are very close to the values of $1/3$ and $2/3$ predicted by a SU(3)$_1$ WZNW model.

\section{Conclusions\label{s4_concl}}
We have provided strong numerical evidence of the criticality of the AFM SU(3) spin chain. Also, we have confirmed that the conformal field theory describing the  chain is effectively the $\text{SU}(3)_1$ WZNW model, by computating the central charge and  scaling dimensions of the lowest excited states of the model, which turn out to be organized according to the IR of SU(3)$_1$ Kac-Moody algebra.

There are many interesting  generalizations of the above models which deserve further study. In particular, a similar ferromagnetic
spin chain is connected with the non-linear $CP^{2}$ sigma mode at $\theta=\pi$ and might be useful to clarify some
controversial problems of the model.  Another interesting problem is to consider larger $SU(N)$ symmetry groups.
In two-dimensional chains, the vacuum state is of N\'{e}el type for $N\leq 4$ and of Spin-Peierls type for $N\geq 5$ 
\cite{harada}. The analysis by means of DMRG technique might shed some light on the transition mechanism.

\begin{acknowledgments}
We would like to thank G. Morandi, C. Degli Esposti Boschi, M. Roncaglia and L. Campos Venuti
for interesting and helpful discussions. One of the authors S.P. would also like to thank
S. Rachel, R. Thomale and A. L\"{a}uchli for very constructive discussions. The work of M.A.  was partially supported by a cooperation
grant INFN-CICYT, the Spanish CICYT grant FPA2006-2315  and  DGIID-DGA (grant2006-E24/2).
\end{acknowledgments}


\begin{thebibliography}{99}

\bibitem{tokura} Y. Tokura, N. Nagaosa, Science {\bf 288}  (2000) 462

\bibitem{greiter} M. Greiter, S. Rachel and D. Schuricht, Phys. Rev. 
{\bf B 75},  (2007) 060401; D. Schuricht and M. Greiter, Europhys. Lett. {\bf 71}  (2005) 987

\bibitem{chen05} S. Chen, C. Wu, S. C. Zhang and Y. Wang, Phys.  Rev.  {\bf B 72}  (2005) 214428

\bibitem{chen06} S. Chen, Y. Wang, W. Q. Ning, C. Wu and H. Q. Lin, Phys. Rev.  {\bf B 74}  (2006) 174424

\bibitem{orignac} R. Citro, E. Orignac, N. Andrei, C. Itoi, S. Qin, J. Phys. : Condens. Matter {\bf 12} (2000)   3041

\bibitem{zarembo} J. A. Minahan and K. Zarembo, JHEP {\bf 0303} (2003) 013

\bibitem{beisert} N Beisert, M Staudacher,  Nucl. Phys. {\bf B 670}(2003) 439

\bibitem{suth} B. Sutherland, Phys. Rev.  {\bf B 12}  (1975) 3795

\bibitem{lai_sut} J. K. Lai, J. Math. Phys. {\bf 15}  (1974) 1675

\bibitem{chang_affl} K. Chang, I. Affleck, G. W. Hayden and Z. G. Soos, J. Phys. : Condens. Matter {\bf 1},   (1989) 153

\bibitem{affl86}I. Affleck, Nucl. Phys.  {\bf B 265} (1986)  409

\bibitem{affl88}I. Affleck, Nucl. Phys.  {\bf B 305}  (1988) 582

\bibitem{itoi} C. Itoi and M. Kato, Phys. Rev. B {\bf 55} (1997) 8295
 
\bibitem{fue} M. Fueheringer, S. Rachel, R. Thomale, M. Greiter and P. Schmitteckert, arXiv:08062563

\bibitem{schmitt} A. Schmitt, K-H M\"{u}tter and M. Karbach, J. Phys  {\bf A 29}  (1996) 3951

\bibitem{diFrancesco} P. Di Francesco, P. Mathieu, D. S\'en\'echal,
{\it Conformal Field Theory}, Springer (1997)

\bibitem{Mukunda} S. Chaturvedi, N. Mukunda, J. Math. Phys. {\bf 43}  (2002) 5262

\bibitem{hakobyan} T. Hakobyan, Nucl. Phys.  \textbf{ B 699}  (2004) 575

\bibitem{holzhey}C. Holzhey, F. Larsen and F. Wilczek, Nucl. Phys. B {\bf 424} (1994) 443

\bibitem{calabrese} P. Calabrese and J. Cardy, J. Stat. Mech. {\bf 0406} (2004) 002  

\bibitem{uimin} G. V. Uimin, JETP Lett {\bf 12} (1970) 225 

%\bibitem{cardy} J.L. Cardy, J. Phys. A: Math. Gen. {\bf 19} (1986) L1093

\bibitem{harada} K. Harada, N. Kawashima and M. Troyer, Phys. Rev. Lett. {\bf 90} (2003) 117203  


\end{thebibliography}
\end{document}